\documentclass[prd,twocolumn,showpacs,preprintnumbers,amsmath,nofootinbib,amssymb,floatfix]{revtex4-1}
\usepackage{graphicx}
\usepackage[sort&compress]{natbib}
\usepackage{subfigure}
\usepackage{amsmath}
\usepackage{amsfonts}
\usepackage{cancel}
\usepackage{amssymb}
\begin{document}
\arraycolsep1.5pt
\newcommand{\Ima}{\textrm{Im}}
\newcommand{\Rea}{\textrm{Re}}
\newcommand{\mev}{\textrm{ MeV}}
\newcommand{\be}{\begin{equation}}
\newcommand{\ee}{\end{equation}}
\newcommand{\ba}{\begin{eqnarray}}
\newcommand{\ea}{\end{eqnarray}}
\newcommand{\gev}{\textrm{ GeV}}
\newcommand{\nn}{{\nonumber}}
\newcommand{\dtres}{d^{\hspace{0.1mm} 3}\hspace{-0.5mm}}
\newcommand{\rts}{ \sqrt s}
\newcommand{\non}{\nonumber \\[2mm]}
\def\J{{J/\Psi}}
\newcommand{\mi}{M_{\pi^0 \eta}}

\title{
Isospin violation in $J/\Psi\to \phi \pi^0 \eta$ decay and the $f_0-a_0$
mixing }

\author{L. Roca}

\affiliation{
Departamento de F\'{\i}sica. Universidad de Murcia. E-30100 Murcia, Spain.\\
}

\date{\today}

\begin{abstract}

The isospin violating  $J/\Psi\to \phi \pi^0 \eta$ decay is thought to
be dominated by the mixing of the $f_0(980)$ and $a_0(980)$ scalar
resonances. We make a theoretical evaluation of the
$J/\Psi\to \phi \pi^0 \eta$ decay extending our own previous model
for other $J/\Psi$ decays into one vector meson and two pseudoscalars
using the techniques of the chiral unitary approach. The scalar
resonances are dynamically generated through the final state
interaction of the pseudoscalar mesons implementing unitarity from the
lowest order ChPT amplitudes. Besides the direct 
$J/\Psi\phi PP$ vertex,
other mechanisms like the sequential exchange of vector and
axial-vector mesons are shown to be important in order to obtain the
actual strength of the mixing.
We get a very good agreement with the $\pi^0 \eta$ 
invariant mass distribution and branching ratio with recent BESIII data.  
Quantification of the $f_0(980)-a_0(980)$ mixing is done and compared
with results from several experiments.

\end{abstract}

\maketitle

\section{Introduction}
\label{Intro}
  
The nature of the scalar mesons has been the
subject of controversial debate for more than thirty years and a final
consensus has not been reached so far. 
The problem stems from the fact that it is very difficult to accommodate
them within the $q\bar q$ picture \cite{Jaffe:1976ig}.
Inspired in the four quarks interpretation \cite{Close:2002zu,Weinstein:1983gd,Vijande:2002rz}
a second kind of molecule picture was developed, in which only mesonic
degrees of freedom were considered \cite{Lohse:1990ew}. In the past
decade new light has been shed by the chiral unitary coupled channel 
approach \cite{Oller:1998hw,Oller:1998zr,Oller:2000ma} which successfully
describes meson-meson interaction in coupled channels up to energies far
beyond the limit of applicability of standard chiral perturbation theory
(ChPT). With the only input of the lowest order ChPT Lagrangians, the
implementation of unitarity in coupled channels and the exploitation
of the analytic properties of the scattering amplitudes, many hadronic
resonances appear dynamically as poles of the
unitarized scattering amplitudes, {\it e.g.}
\cite{Kaiser:1995eg,angels,ollerulf,Lutz:2001yb,Oset:2001cn,Hyodo:2002pk,Jido:2003cb,Roca:2005nm,Borasoy:2005ie,Oller:2006jw}.
 These are usually called
molecules or dynamically
generated resonances, like the $f_0(980)$ and $a_0(980)$
\cite{Oller:1997ti,Pelaez:2003dy,Kaiser:1998fi,Markushin:2000fa,Nieves:1999bx},
of interest for the present work.

One interesting issue of the $a_0$ and $f_0$ resonances is that
they can mix 
 violating isospin, as it was first
suggested in ref.~\cite{Achasov:1979xc}. This mixing is expected to 
help clarify the nature of these resonances. 
It is thought~\cite{Achasov:1979xc,Hanhart:2003pg} to be 
enhanced and only relevant in the narrow $8 \mev$ region between
the $K^+K^-$ and $K^0\bar K^0$ thresholds, ($987.4\mev$ and $995.2\mev$ 
respectively).
Several reactions could be sensitive to the  $f_0-a_0$ mixing, for
example $pn\to d \eta\pi^0$~\cite{Kudryavtsev:2001ee},
$\gamma p\to p\pi^0\eta$~\cite{Kerbikov:2000pu} or
$\pi^-p\to\pi^0\eta n$~\cite{Achasov:2003se}. 
Very recently the isospin violation in $\eta(1405)$ decays 
into $f_0(980)\pi^0$ have been experimentally measured \cite{BESIII:2012aa}
and theoretically studied in a simultaneous
 work~\cite{Aceti:2012dj} to the present one and using also techniques
 of the chiral unitary approach.

Until very recently \cite{Ablikim:2010aa} there has been
no conclusive experimental
information on this mixing. In ref.~\cite{Ablikim:2010aa} the 
$\pi^0 \eta$ invariant
mass distribution of the $J/\Psi\to \phi \pi^0 \eta$ decay
 has been measured experimentally 
by the BESIII collaboration  in Beijing and a clear peak in
the window between the $K^+K^-$ and $K^0\bar K^0$ thresholds has been
observed. The branching ratio of the 
$J/\Psi\to \phi f_0(980)\to\phi a^0_0(980)\to \phi \pi^0 \eta$ 
is determined in  \cite{Ablikim:2010aa} to be of the order of $3.3\times
10^{-6}$ with a significance of about $3.4\sigma$ which represents the
most significant evidence of the isospin violating $f_0-a_0$ mixing to
date.
From the theoretical side a first estimation of the mixing in the
$J/\Psi\to \phi \pi^0 \eta$ decay was done in ref.~\cite{Wu:2007jh}
using a simple  model with explicit scalar resonance propagators coupling
to the pseudoscalar pairs. In ref.~\cite{Hanhart:2007bd} the 
$J/\Psi\to \phi \pi^0 \eta$ was calculated from the implementation of
the final rescattering of the two pseudoscalars in the direct
$J/\Psi\phi PP$ vertex using the chiral unitary approach.

In the work of ref.~\cite{Roca:2004uc} we developed a very detailed
and thorough
model (one of the most comprehensive to date)
for the decay of the $J/\Psi$ into a vector meson and two
pseudoscalars, in particular for the isospin conserving decays into
$\omega\pi^+\pi^-$, $\phi\pi^+\pi^-$, $\omega K^+ K^-$ and $\phi
K^+K^-$. Besides the direct $J/\Psi\phi PP$ vertex and the
implementation of the final state interaction of the two pseudoscalars
 other new mechanisms were considered in ref.~\cite{Roca:2004uc} where 
the $\J$ decays into a vector  or axial-vector and a pseudoscalar meson
and the vector or axial-vector subsequently decays into the final vector
and another pseudoscalar. We call these mechanisms sequential vector and
axial-vector exchange. In this sequential mechanisms also the
final pseudoscalar-pseudoscalar interaction is 
implemented from which the scalar resonances
are naturally generated dynamically. We found a very good
reproduction of the
experimental mass distribution
also in the $\sigma(500)$ and $f_0(980)$ regions, not trivial at all.

 In view of the power
of the model of ref.~\cite{Roca:2004uc}, in the present work we are
going to extend it to the $J/\Psi\to \phi \pi^0 \eta$ decay in order to
analyze the $f_0-a_0$ mixing. The isospin violation will be implemented
essentially
from leading order quark mass differences and electromagnetism,
similarly as in ref.~\cite{Hanhart:2007bd}. In practice it will come
essentially from the difference between the masses of the 
charged and neutral kaons
and pions in the loops and the meson-meson scattering amplitudes.
One of the main differences with ref.~\cite{Hanhart:2007bd} is the new
sequential mechanisms of ref.~\cite{Roca:2004uc} which, advancing some
results, turn out to be important due to strong interferences with the
dominant loops from direct production.


\section{The $f_0-a_0$ mixing through meson-meson
 unitarization}

In the literature several unitarization procedures have been used to obtain a
scattering matrix fulfilling exact unitarity in coupled channels, like the Inverse
Amplitude Method \cite{dobado-pelaez,Oller:1998hw}
 or the N/D method \cite{Oller:1998zr}. 
In this latter work the equivalence
with the Bethe-Salpeter  equation used in \cite{Oller:1997ti}
 was established.
In the present work we use the Bethe-Salpeter  equation
which leads to the following unitarized amplitude in coupled
channels
\be
t=[1-VG]^{-1}V.
\label{eq:BS}
\ee

Diagrammatically Eq.~(\ref{eq:BS}) 
is equivalent to resum
 the series
expressed by the thick dot in Fig.~\ref{fig:chiral_loops}. 
In Eq.~(\ref{eq:BS}) $G$ is a diagonal matrix
 with the $l-$th element, $G_l$, being the two meson
loop function containing the two pseudoscalar mesons of the
$l-$th channel:
\begin{equation}
G_{l}(\sqrt{s})= i \, \int \frac{d^4 q}{(2 \pi)^4} \,
\frac{1}{(P-q)^2 - M_l^2 + i \epsilon} \,
 \frac{1}{q^2 - m^2_l + i
\epsilon}, \label{loop}
\end{equation}
\noindent with $P$ the total incident momentum, which in the center
of mass frame is $(\sqrt{s},0,0,0)$.
The $G_l$ integral above is divergent, and hence it has to be
regularized, which can be done  either with a three momentum
cutoff, $q_\textrm{max}$, or
with dimensional regularization.
 The connection between  both methods was
shown in Refs.~\cite{ollerulf,Oller:1998hw}. We use the cutoff
regularization since this was the one used in 
ref.~\cite{Roca:2004uc} with a  value $q_\textrm{max}=1\gev$
which satisfactorily describes a wide phenomenology for
meson-meson interaction \cite{Oller:1997ti}.

In Eq.~(\ref{eq:BS}) $V$, the kernel of the BS equation,
is a matrix containing the s-wave projected
scattering amplitude for the
two pseudoscalar mesons. 
Since the main source of isospin violation is the different mass
between the different charge states, specially the kaons, it is
much more convenient to work in the charge, or particle, basis
instead of the isospin basis used 
 in  ref.~\cite{Roca:2004uc} and in typical  chiral
unitary approach works.
Thus the different channels considered 
are $K^+ K^-$, $K^0 \bar K^0$,
$ \pi^+ \pi^-$, $ \pi^0 \pi^0$, $ \eta \eta$ and $ \pi^0 \eta$.

Like in ref.~\cite{Hanhart:2007bd}  
the tree level vertices, $V$, in Eq.~(\ref{eq:BS}) are
obtained from the lowest order chiral Lagrangian with
isospin-breaking/electromagnetic effects \cite{Urech:1994hd},
from where the potential between the different channels can be
evaluated. The explicit expressions can be found in Eq.~(D2)
 of ref.~\cite{Hanhart:2007bd} but all the amplitudes have a
 difference sign with respect to our notation and the
  amplitudes
 involving a $\pi^0 \pi^0$ or $\eta\eta$ must be multiplied by
 an $1/\sqrt{2}$ factor to agree with the unitary normalization
 of ref.~\cite{Oller:1997ti}.
 As explained in ref.~\cite{Hanhart:2007bd}, 
 these elementary meson-meson amplitudes  express the isospin
 breaking effects in terms of the $\pi^0\eta$ mixing angle
 $\epsilon$, and the charge to neutral pion and kaon mass
 difference, $\Delta \pi=m_{\pi^+}^2-m_{\pi^0}^2$ 
 and $\Delta K=m_{K^+}^2-m_{K^0}^2$ respectively.

The use of these tree level amplitudes and the different 
masses between different charge states in the two meson loop
functions is what generates the isospin violation in the
unitarized amplitudes.

\begin{figure}[!h]
\begin{center}
\includegraphics[width=0.9\linewidth]{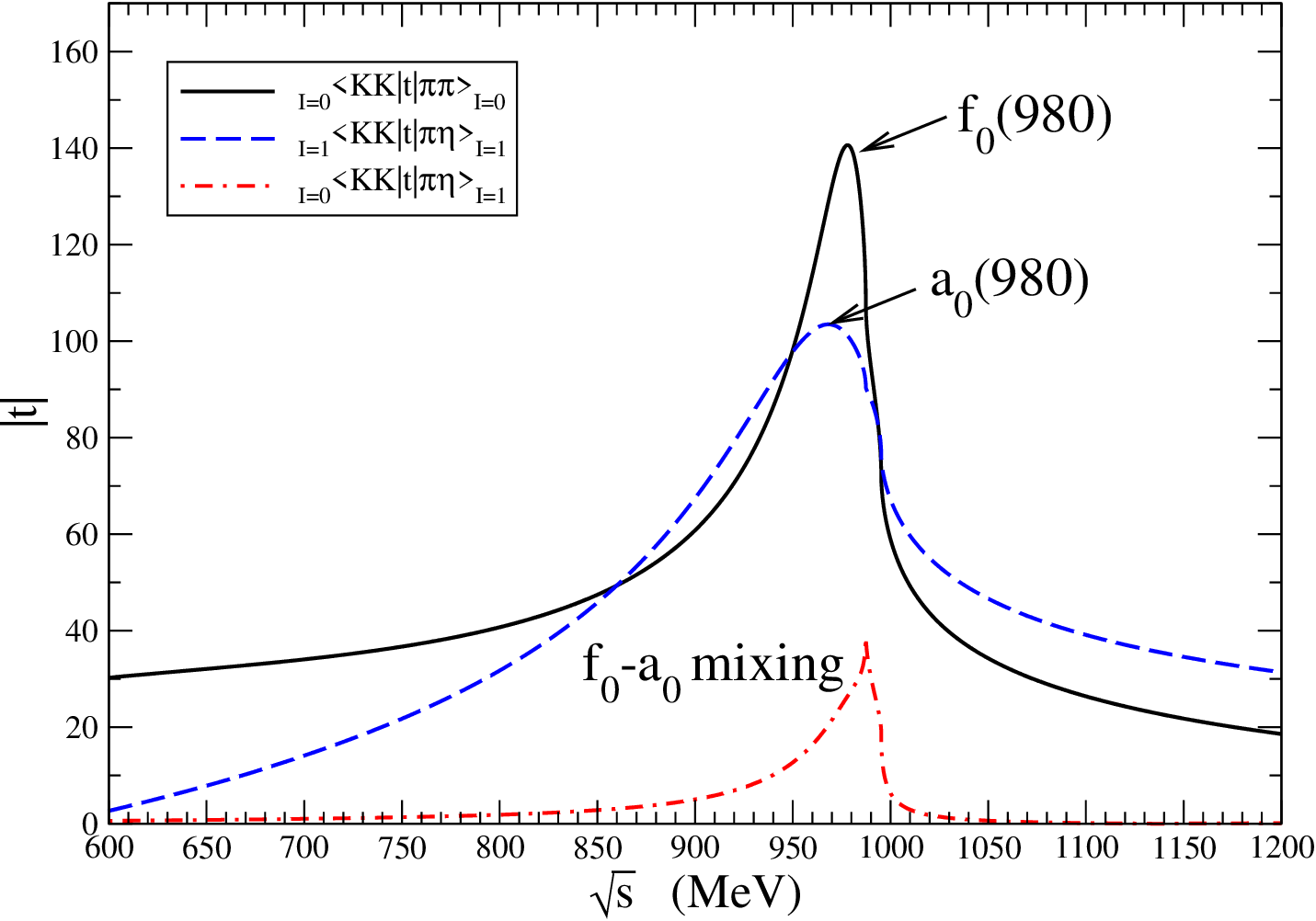}
\caption{Modulus of the  isospin conserving 
amplitudes $_{I=0}\left<K\bar K|t|\pi\pi\right>_{I=0}$,
$_{I=1}\left<K\bar K|t|\pi\eta\right>_{I=1}$ 
and the isospin violating $_{I=0}\left<K\bar K|t|\pi\eta\right>_{I=1}$.}
\label{fig:t_MMMM}
\end{center}
\end{figure}

For an illustration of this effect we show  in 
Fig.~\ref{fig:t_MMMM} the
modulus  of the 
amplitudes $_{I=0}<K\bar K|t|\pi\pi>_{I=0}$
$_{I=1}<K\bar K|t|\pi\eta>_{I=1}$ 
and $_{I=0}<K\bar K|t|\pi\eta>_{I=1}$.
In the conserving isospin amplitudes we see clear peaks
corresponding to the $f_0$ and $a_0$ resonances 
 for the isospin $I=0$ and
$I=1$ respectively which are dynamically generated.
The  $_{I=0}<K\bar K|t|\pi\eta>_{I=1}$ amplitude manifests the
 $f_0-a_0$ mixing dynamically obtained within the chiral unitary
 approach.
 Note the difference between the
strength of the isospin violating amplitude with respect to 
the conserving ones
 which gives an idea of the size of the isospin violation
 through $f_0-a_0$ mixing.  

\section{The $J/\Psi\to \phi \pi^0 \eta$ model within the chiral unitary
approach}
\label{sec2}

In this section we explain the model we use for the evaluation of the 
$J/\Psi\to \phi \pi^0 \eta$, which is based in our previous work of
ref.~\cite{Roca:2004uc}, and how to implement into it the isospin
violation explained in the previous section. Our model of
ref.~\cite{Roca:2004uc} successfully identified and established  the
relevant mechanisms that contribute to the decay of the $J/\Psi$ into
one vector meson and two pseudoscalars. In particular,
ref.~\cite{Roca:2004uc} focused on the isospin conserving channels
$\omega\pi\pi$, $\phi\pi\pi$, $\omega K \bar{K}$ and $\phi K\bar{K}$.
The main contribution was found to be the mechanisms containing  the
''direct'' $\J V PP$ vertex implementing also the final state
interaction of the pseudoscalar pair using the techniques of the chiral
unitary approach from where the scalar mesons appear dynamically in the
meson-meson scattering amplitudes without the need to include them as an
explicit degree of freedom. The actual shape of the amplitudes in the
complex plane, and in particular in the physical real axis, couplings of
the scalars to the different channels, etc, come naturally with the only
input of the lowest order chiral Lagrangian and the implementation of
unitarity in coupled channels. Only one free parameter (for the
regularization of the two meson loop function) is fitted to the
experimental meson-meson data. (See for instance
refs.~\cite{Oller:1997ti,Oller:2000ma} and  the references cited in the
introduction). The model also includes other mechanisms where the $\J$
decays into a vector and or axial-vector and a pseudoscalar meson and
the vector subsequently decays into the final vector an the other final
pseudoscalar. Also the final state interaction is implemented as before.
These new mechanism where found to be crucial in order to obtain the
actual shape and strength of the mass distributions in the different
channels because, in spite that they are small by themselves the interference
with the dominant mechanism is very important.

 In this section we extend and adapt the model of ref.~\cite{Roca:2004uc}  to the $J/\Psi\to \phi \pi^0 \eta$
stressing the differences of this channel with those studied in 
ref.~\cite{Roca:2004uc}. We refer the reader to \cite{Roca:2004uc} for the 
details and here we only address the particularities,
differences and
the points   of special interest
for the present work.

\begin{figure}[!h]
\begin{center}
\includegraphics[width=0.7\linewidth]{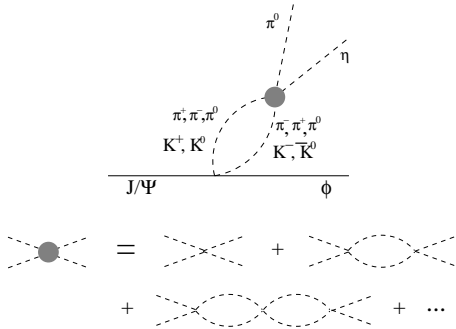}
\caption{Diagrams for the final state interaction from the direct $\J \phi PP$
vertex implementing the final state interaction represented by the solid circle.}
\label{fig:chiral_loops}
\end{center}
\end{figure}

\begin{figure}[!h]
\begin{center}
\includegraphics[width=0.95\linewidth]{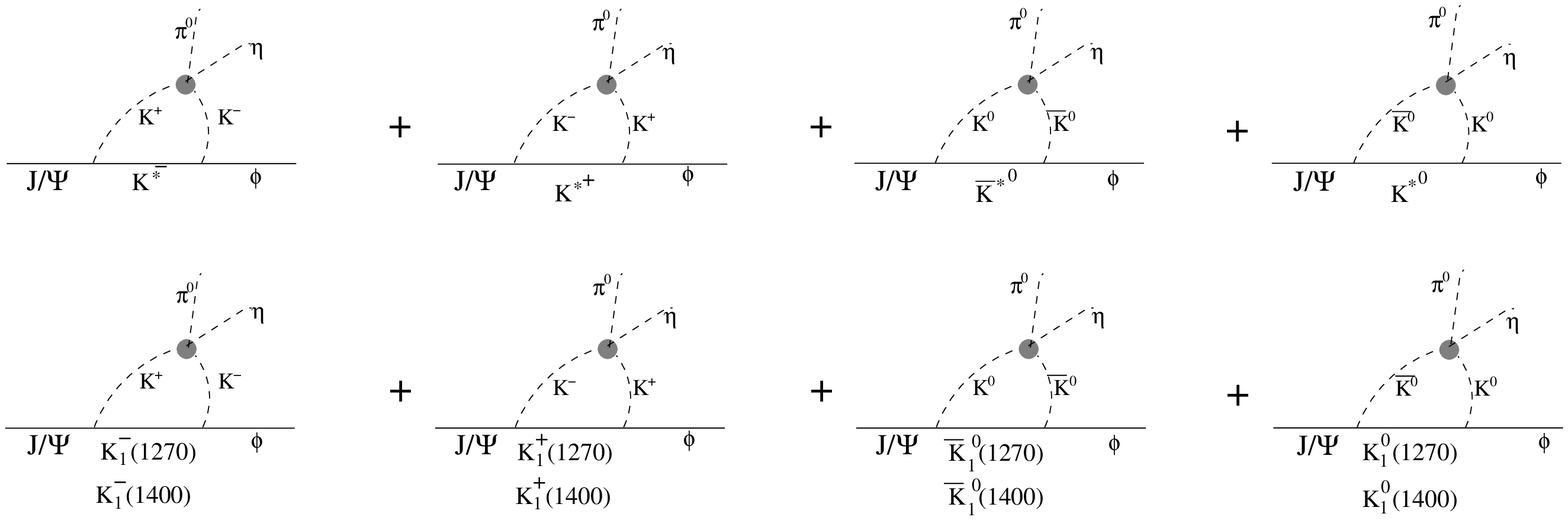}
\caption{Diagrams for the sequential exchange of vector ($K^*(892)$) and
axial-vectors ($K_1(1270)$ and $K_1(1400)$) with final state interaction.}
\label{fig:loops_sequential}
\end{center}
\end{figure}

The  $J/\Psi\to \phi \pi^0 \eta$ decay can proceed only through processes
involving isospin violation.  We do not consider tree level $J/\Psi \phi \pi^0 \eta$
vertex since in ref.~\cite{Hanhart:2007bd} was shown that
isospin violation in the production operator is very small, of the order of $4\%$.
Other possible source of isospin violation like the
soft photon exchange in the meson propagation was also found to be small in 
ref.~\cite{Hanhart:2007bd}. 
Thus the main contribution to isospin violation
comes from the $f_0-a_0$ mixing which in our formalism is generated essentially 
from the loops and potentials implicit in the 
meson-meson
 unitarized scattering
amplitudes, thick dots of 
Figs.~\ref{fig:chiral_loops} and \ref{fig:loops_sequential}.

For the evaluation of the diagrams of Fig.~\ref{fig:chiral_loops}, which we call
"direct" production in the following,  we need first  the  $\J \phi PP$
vertex which in ref.~\cite{Roca:2004uc} it was constructed using $SU(3)$ arguments
to relate the different channels and also parametrizing the amplitudes in a way
which manifest the OZI rule violation, in a similar way as 
in ref.~\cite{Meissner:2000bc}. This introduces two a priori unknown constants: an
overall coupling $\tilde g$ and the OZI violation parameter, $\lambda_\phi$.
These free parameters were fitted in ref.~\cite{Roca:2004uc} to the invariant mass
distributions of the $\J\to\omega\pi^+\pi^-$ and $\J\to\phi\pi^+\pi^-$.

Following the procedure of ref.~\cite{Roca:2004uc} we obtain

\ba
\nonumber
t_{\J \phi
K^+K^-}&=&-\frac{\tilde{g}}{3}(2\nu+1)\,\epsilon^*\cdot\epsilon \\ \nonumber
t_{\J \phi\pi^+\pi^-}&=&
t_{\J \phi\pi^0\pi^0}
-\frac{2\tilde{g}}{3}(\nu-1)\,\epsilon^*\cdot\epsilon \label{eq:tvertexOll}
\ea
with $\nu=\frac{\sqrt{2}+2\lambda_\phi}{\sqrt{2}-\lambda_\phi}$.

After the implementation of the rescattering of the 
final pseudoscalar pair, the amplitude for the direct production is given
by\footnote{In Eq.~(\ref{eq:tchiral_loops}) the amplitudes
involving $\pi^0\pi^0$ or $\eta\eta$ are in {\it good} normalization,
not in the unitary normalization used in the Bethe-Salpeter
equation,  Eq.~(\ref{eq:BS}).}
 
 \ba
\nonumber
&&t_{\J\to\phi\pi^0\eta}=-\widetilde{g}\epsilon^*·\epsilon\times\nonumber\\
\times&& \bigg[\frac{1+2\nu}{3}
       (G_{K^+K^-}t_{K^+K^-\to\pi^0\eta}+G_{K^0\bar K^0}t_{K^0\bar K^0\to\pi^0\eta})\nonumber\\
      &&+\frac{\nu-1}{3}
      (2G_{\pi^+\pi^-}t_{\pi^+\pi^-\to\pi^0\eta}
      +G_{\pi^0\pi^0}t_{\pi^0\pi^0\to\pi^0\eta})\nonumber\\
      &&
      + \frac{2}{9}(2+\nu)G_{\eta\eta}t_{\eta\eta\to\pi^0\eta}
      \bigg]
\label{eq:tchiral_loops}
\ea

The mechanisms with sequential exchange of vectors and axial-vector mesons
for the $J/\Psi\to \phi \pi^0 \eta$ decay are shown in
Fig.~\ref{fig:loops_sequential}.  For this
particular decay the vector exchanged is a $K^*(892)$ and the axial-vectors
are
the $K_1(1270)$ and $K_1(1400)$. 
We need the vertices $\J$-vector-pseudoscalar ($\J VP$),
vector-vector-pseudoscalar ($VVP$), $\J$-axial--vector-pseudoscalar
($\J AP$) and axial--vector-vector-pseudoscalar ($AVP$) which we
obtain from the Lagrangians of refs.~\cite{Roca:2004uc}, 
\cite{escribano}, \cite{Roca:2004uc} and \cite{axials} respectively.
After implementing the final state interaction of the pseudoscalar pair the
amplitudes for the sequential exchange of the $K^*$ meson is
\begin{equation}
t^{\mu\nu} =G \overline{G}
\, \bar{t}'^{\mu\nu} \, ( t_{K^+K^-,\pi^0\eta}+t_{K^0\bar K^0,\pi^0\eta})
\end{equation}
where $G$ is the coupling in  the $VVP$ Lagrangian, 
$\overline{G}$ the coupling in the $\J VP$ one, and $\bar{t}'^{\mu\nu}$
contains the three meson loop function, $K^*K\bar K$, including the momentum structure of
the different vertices. (See ref.~\cite{Roca:2004uc} for details and values
of the constants\footnote{In ref.~\cite{Roca:2004uc} two different results for
the fit were obtained depending on the sign of $\overline{G}$.
Our result in \cite{Roca:2004uc} for $\overline{G}>0$ is 
compatible with the later work of ref.~\cite{Lahde:2006wr} and thus 
we use
this result of the fit of \cite{Roca:2004uc} for 
the evaluation of our central results. Our
 other result with $\overline{G}<0$ will be included in the error
 uncertainties.}).

For the $K_1(1270)$ axial-vector exchange we get

\ba
t^{\mu\nu} &=& \frac{8}{M_\J m_\phi m_{K_1(1270)}^2}
c\overline{D}(cD+sF)\nn\\
&\times&\left[(q_\J\cdot q_\phi g^{\mu\nu}-q_\phi^{\mu}\cdot q_\J^{\nu}) G_{KK}
-\tilde{t}'^{\mu\nu}\right] \nn\\
&\times&( t_{K^+K^-,\pi^0\eta}+t_{K^0\bar K^0,\pi^0\eta})
\label{eq:cc}
\ea
where $D$ and $F$ are couplings in the  $AVP$ Lagrangian, $\overline{D}$
is the coupling of the $\J AP$, $s$ and $c$ parameterize the mixing
 between the isospin $1/2$ members of the axial-vector
octets to give the physical $K_1(1270)$ and $K_1(1400)$ states and $\tilde{t}'^{\mu\nu}$
contains the three meson loop function, $K_1(1270)K\bar K$, including 
the momentum structure of
the different vertices \cite{Roca:2004uc}. For the diagrams with
$K_1(1400)$ intermediate state 
 the amplitude is the same but changing
$m_{K_1(1270)}\to m_{K_1(1400)}$, $F\to -F$, $c\to s$ and $s\to
c$ and replacing the masses and widths of the $K_1(1270)$
by those of the $K_1(1400)$ in the evaluation of
$\tilde{t}'^{\mu\nu}$.
 
\section{Results}

The main observable that we are going to evaluate is
the differential
decay width of the $\J$ with respect to the $\pi^0 \eta$
invariant mass, $M_{\pi^0 \eta}$, which in the $\J$ rest frame 
can be evaluated as

\begin{equation}
\frac{d\Gamma}{dM_{\pi^0 \eta}}=\frac{\mi}{64\pi^3m_\J^2}
\int_{m_{\pi^0}}^{m_\J-\omega_{\phi}-m_\eta}d\omega_{\pi^0}
\overline{\sum}|t|^2\Theta(1-\cos\bar{\theta}^2)
\end{equation}
where  $\Theta$ is the step function,  
$\cos\bar{\theta}=
((m_\J-\omega_\phi-\omega_{\pi^0})^2-m_\eta^2-
|\vec{q}_\phi|^2-|\vec{p}_{\pi^0}|^2)/(2|\vec{q}_\phi|
 |\vec{p}_{\pi^0}|)$, where
$\bar{\theta}$ is the angle between $\vec{p}_{\pi^0}$
 and $\vec{q}_\phi$ and $\omega_i=\sqrt{\vec q_i\,^2+m_i^2}$.
The final polarization sum and initial average is
\be
\overline{\sum}|t|^2=\frac{1}{3}\sum_{\mu\mu'\nu\nu'}
\left(-g^{\mu{\mu}'}+\frac{q_\J^\mu q_\J^{\mu'}}{m_\J^2}\right)
\left(-g^{\nu{\nu}'}+\frac{q_\phi^\nu q_\phi^{\nu'}}{m_\phi^2}\right)
t_{\mu\nu}{t^*}_{\mu'\nu'}
\ee
To obtain the total scattering amplitude, 
$t\equiv\epsilon^*_\mu\epsilon_\nu t^{\mu\nu}$, all 
the mechanisms
explained in the previous section must be added.

\begin{figure}[!h]
\begin{center}
\includegraphics[width=0.9\linewidth]{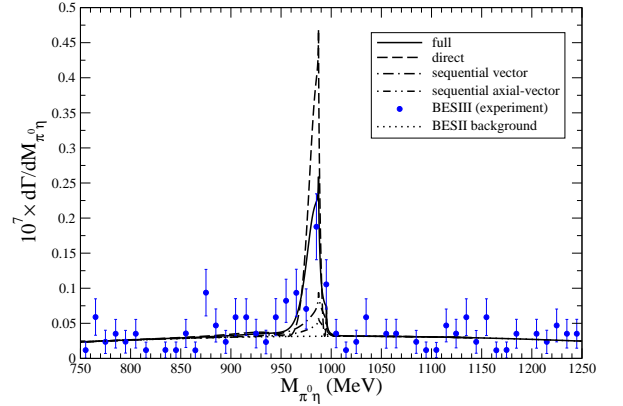}
\caption{Different contributions to the $\pi^0 \eta$ mass distribution
 in the $J/\Psi\to \phi \pi^0 \eta$ decay. In the theoretical
 calculation we have added the experimental background, dotted line, as
 given in \cite{Ablikim:2010aa} to ease the comparison with the data.  }
\label{fig:inv_mass_contrib_zoom_dentro}
\end{center}
\end{figure}

\begin{figure}[!h]
\begin{center}
\includegraphics[width=0.9\linewidth]{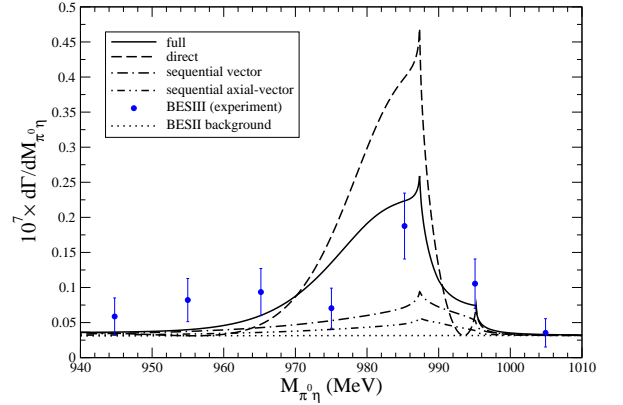}
\caption{Same as Fig.~\ref{fig:inv_mass_contrib_zoom_dentro}
but zoomed around the peak region}
\label{fig:inv_mass_contrib_zoom}
\end{center}
\end{figure}
\begin{figure}[!h]
\begin{center}
\includegraphics[width=0.9\linewidth]
{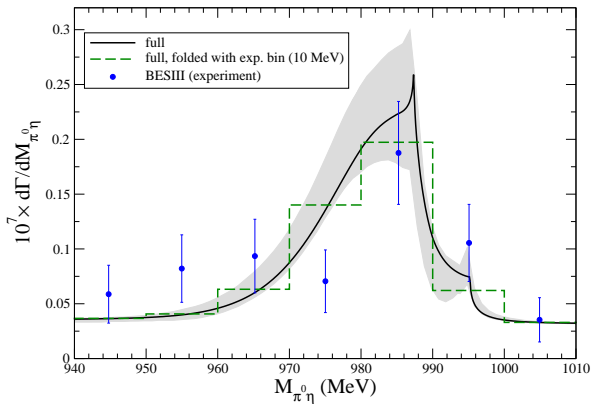}
\caption{Solid line and error band: 
same as in fig.~\ref{fig:inv_mass_contrib_zoom}.
Dashed line: full result plus experimental background 
implementing a folding with the experimental
resolution, (bins $10\mev$ wide).}
\label{fig:inv_mass_contrib_zoom_error_bin}
\end{center}
\end{figure}
In figs.~\ref{fig:inv_mass_contrib_zoom_dentro} and
 \ref{fig:inv_mass_contrib_zoom}
we show the contribution of the different mechanisms to the 
 $\pi^0 \eta$ mass distribution, 
$\frac{d\Gamma}{dM_{\pi^0 \eta}}$.
(Fig.~\ref{fig:inv_mass_contrib_zoom} is just a zoom
of Fig.~\ref{fig:inv_mass_contrib_zoom_dentro}
around the peak region).
 The experimental data is the result
from BESIII~\cite{Ablikim:2010aa}. The dotted line represents
the background given
in the experiment \cite{Ablikim:2010aa}. Since it is not subtracted in
the original experimental data, we have added this
background to our theoretical results shown in the figure for
the sake of comparison  with the BESIII data. The data provided in  
ref.~\cite{Ablikim:2010aa} is given in number of events but, since 
 ref.~\cite{Ablikim:2010aa} also provides the total number of $\J$
  produced
 and detector efficiencies, it is possible to obtain the absolute
 normalization for  the observable shown in the figure.
The solid line represents our full theoretical result. The dashed line
represent the contribution of the direct $\J V PP$ mechanism, 
fig.~\ref{fig:chiral_loops}. The dashed-dotted line is the result from
the mechanisms with sequential exchange of the vector meson, upper row
in fig.~\ref{fig:loops_sequential}, and the dashed-double--dotted line
is
the result from the sequential exchange of axial-vectors, bottom row 
in fig.~\ref{fig:loops_sequential}.
 We see that the loops from direct mechanisms by themselves,
 which is essentially the only mechanism considered in ref.~\cite{Hanhart:2007bd}
 overestimate the strength in about a factor two with respect to the
 experimental data at the peak.
The new mechanisms
 containing the sequential exchange of the vector meson ($K^*$) and 
 the axial-vectors, ($K_1(1270)$ and $K_1(1400)$),
are not very large by themselves but they are crucial in order to provide
the right strength of the distribution when added coherently to the
direct mechanisms. Thus these sequential mechanisms are important
in order to obtain the right $f_0-a_0$ mixing.

In Fig.~\ref{fig:inv_mass_contrib_zoom_error_bin} we show our final
result including the theoretical error band
obtained implementing a Monte Carlo gaussian sampling of the parameters
used in the model within their error bounds. (See  ref.~\cite{Roca:2004uc}
for the values and uncertainties of these parameters). 
To ease the comparison with the experimental
data we also show with dashed lines our full result averaged in bins
of $10\mev$, like the experimental resolution. 
 Note that there is
no fit to this observable. All the parameters of the model
are obtained from other
reactions or theoretical works, thus our results are genuine predictions.
For the branching ratio $BR(J/\Psi\to \phi \pi^0 \eta)$
we obtain $(4.8\pm0.5)\times10^{-6}$ which can be compared to the BESIII
result $(5.0\pm2.7(\textrm{stat})\pm 1.7(\textrm{sys}))\times10^{-6}$
 \cite{Ablikim:2010aa}.

In ref.~\cite{Wu:2008hx} an observable called "mixing intensity'',
$\xi_{fa}$, was defined in order to quantify the effect of the
$f_0(980)\to a_0(980)$ transition independently of the 
production reaction. For the present case it would be defined as

\be
\xi_{fa}(M)=\frac{d\Gamma_{J/\Psi\to \phi f_0\to \phi a_0\to
 \phi \pi^0 \eta}}{d\Gamma_{J/\Psi\to \phi f_0\to 
 \phi \pi \pi}}
 \label{eq:xi1}
\ee
where $M$ is the invariant mass of the two pseudoscalars in the final
state. In our formalism it has no sense to strictly 
isolate the $f_0$ or $a_0$
contributions by themselves since the chiral unitary approach generates
the full amplitude from where the definition of the resonance is
 a matter
of convention. Thus, the meaningful observable is the ratio
\be
\xi'_{fa}(M)=\frac{d\Gamma_{J/\Psi\to \phi \pi^0 \eta}}
{d\Gamma_{J/\Psi\to \phi \pi \pi}}
\ee
which close to the $f_0$ mass region should be very similar to 
Eq.~(\ref{eq:xi1}). The calculation for the $J/\Psi\to \phi \pi \pi$
within our formalism was done in Ref.~\cite{Roca:2004uc}\footnote{
In Ref.~\cite{Roca:2004uc} the  $J/\Psi\to \phi \pi^+ \pi^-$
channel is
provided. One has to take into account that 
$\Gamma_{J/\Psi\to \phi \pi^+ \pi^-}=\frac{2}{3}
\Gamma_{J/\Psi\to \phi \pi \pi}$
 since $2/3$ go to $\pi^+ \pi^-$ and $1/3$ to 
$\pi^0 \pi^0$.}.

\begin{figure}[!h]
\begin{center}
\includegraphics[width=0.9\linewidth]{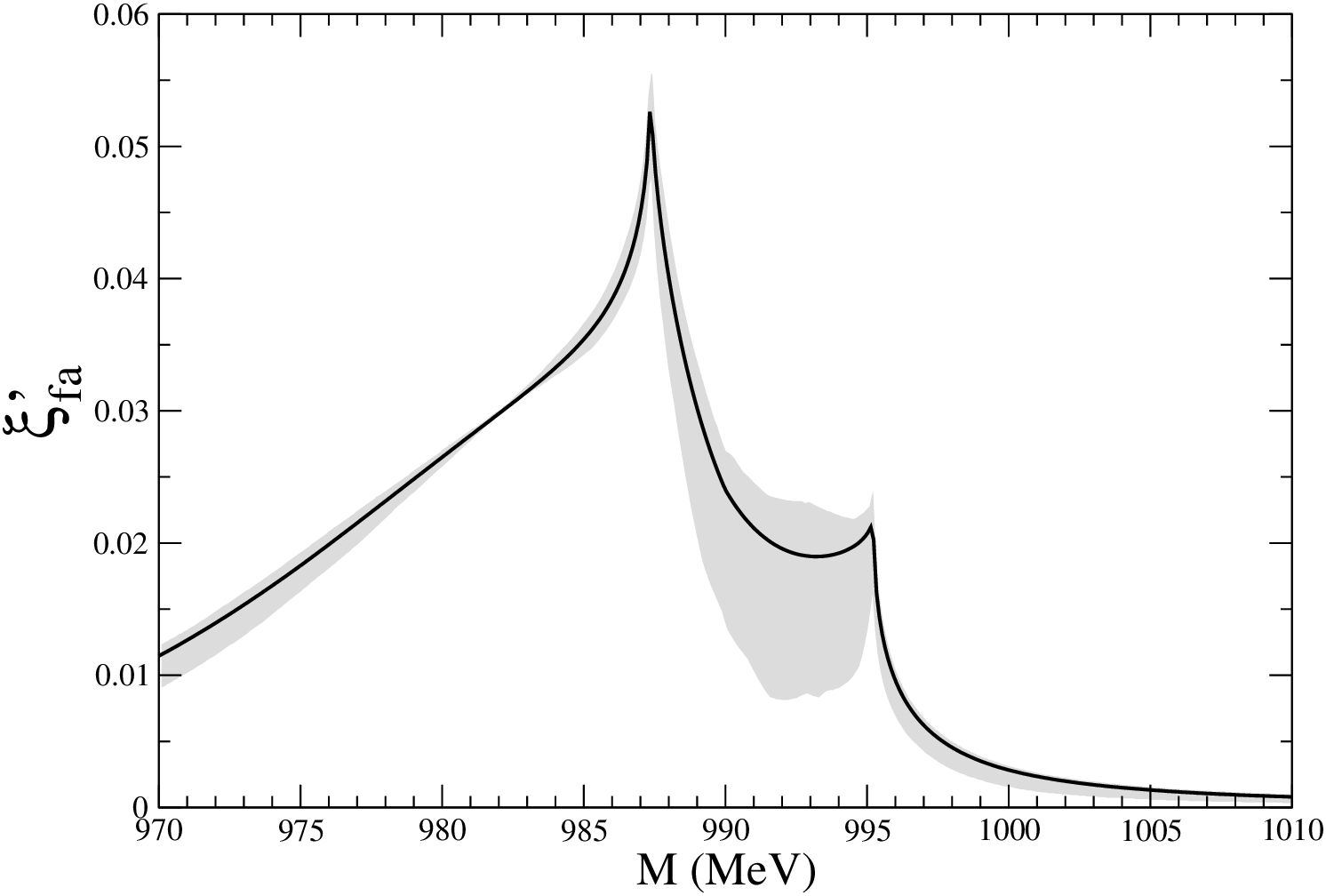}
\caption{$f_0\to a_0$ mixing intensity.}
\label{fig:mixint}
\end{center}
\end{figure}

The result for the mixing intensity is shown in Fig.~\ref{fig:mixint}.
The mixing intensity can also be obtained from other production
processes, like {\it e.g.} radiative $\phi$ decays $\phi\to\pi^0\eta\gamma$ and
 $\phi\to\pi^0\eta\gamma$. In order to compare the 
 mixing intensity with
 experimental data, in ref.~\cite{Wu:2008hx} the mixing intensity at 
 $M=991.4\mev$, which is the two kaon threshold for average kaon masses,
  was obtained for different experiments 
 \cite{Achasov:2000ym,Achasov:2000ku,Aloisio:2002bsa,Aloisio:2002bt,
 Teige:1996fi,Bugg:1994mg}.
From 
 Fig.~\ref{fig:mixint} we obtain $\xi_{fa}(M=991.4\mev)=0.020^{+0.003}_{-0.011}$ to
 be compared with the experimental results, quoted in \cite{Wu:2008hx},
0.088, 0.034, 0.019, 0.027 for SND \cite{Achasov:2000ym,Achasov:2000ku},
KLOE \cite{Aloisio:2002bsa,Aloisio:2002bt},
BNL \cite{Teige:1996fi} and CB \cite{Bugg:1994mg}
respectively.

\section{Summary}
\label{sec4}

We have studied theoretically the isospin violating 
$J/\Psi\to \phi \pi^0 \eta$ extending our previous model
 \cite{Roca:2004uc} on the $J/\Psi\to \phi PP$ decays.
The main contribution of the model are the mechanisms 
containing  the
''direct'' $\J V PP$ vertex implementing also the final state
interaction of the pseudoscalar pair.
This final state interaction is evaluated using the techniques of the
chiral unitary approach, which implements unitarity in coupling channels
with the only input of the lowest order pseudoscalar-pseudoscalar chiral
Lagrangian. At the relevant energies of the present work, 
the scalar $f_0(980)$ and $f_0(980)$ appear naturally (dynamically)
in the unitarized
scattering amplitudes without the need to include them as explicit
degrees of freedom.
Due to the strong interference with the previous mechanisms, 
we find also very relevant for the final results the mechanisms 
where the $\J$
decays into a vector and or axial-vector and a pseudoscalar meson and
the vector subsequently decays into the final vector and the other final
pseudoscalar, where the final state interaction is also implemented.

The isospin violation comes essentially 
from the difference between the masses of the 
charged and neutral kaons
and pions in the loops and the meson-meson scattering amplitudes.

We find a very good agreement with the experimental BESIII results
for the $\pi^0\eta$ mass distribution and branching ratio. It is worth
noting that without the inclusion of the sequential exchange mechanisms
we would have obtained results a factor about 2 larger.

The $f_0-a_0$ mixing intensity, as defined in ref.~\cite{Wu:2008hx}, is
also evaluated from the ratio between the 
$J/\Psi\to \phi \pi^0 \eta$ and $J/\Psi\to \phi \pi \pi$
decays and we obtain a fair agreement with most of the experimental
results evaluated in ref.~\cite{Wu:2008hx}.

\section*{Acknowledgments}
I thank Eulogio Oset for suggesting me to study this topic and for
useful discussions.
 This work is partly supported by DGICYT contracts  FIS2006-03438,
 the Generalitat Valenciana in the program Prometeo 2009/09, MEC  FPA2010-17806, the Fundaci\'on S\'eneca  11871/PI/090 and 
the EU Integrated Infrastructure Initiative Hadron Physics
Project under Grant Agreement n.227431.

\end{document}